\documentclass[twocolumn]{aastex631}

\usepackage{graphicx} 

\usepackage{amsmath}
\usepackage{relsize}

\begin{document}

\title{Scale Filtering Analysis of Kinetic Reconnection and its Associated Turbulence}

\author[0000-0002-0786-7307]{Subash Adhikari}
\affiliation{Department of Physics and Astronomy and the Center for KINETIC Plasma Physics, West Virginia University, Morgantown, West Virginia 
26506, USA}\email{subash.adhikari@mail.wvu.edu}

\author[0000-0003-2965-7906]{Yan Yang}
\affiliation{Department of Physics and Astronomy, University of Delaware, Newark, DE 19716, USA}

\author[0000-0001-7224-6024]{William H. Matthaeus}
\affiliation{Department of Physics and Astronomy, University of Delaware, Newark, DE 19716, USA}

\author[0000-0002-5938-1050]{Paul A. Cassak}
\affiliation{Department of Physics and Astronomy and the Center for KINETIC Plasma Physics, West Virginia University, Morgantown, West Virginia 26506, USA}

\author[0000-0001-7224-6024]{Tulasi N. Parashar}
\affiliation{School of Chemical and Physical Sciences, Victoria University of Wellington, Wellington 6012, New Zealand}

\author[0000-0001-7224-6024]{Michael A. Shay}
\affiliation{Department of Physics and Astronomy, University of Delaware, Newark, DE 19716, USA}

\begin{abstract}
Previously, 
using an incompressible von K\'arm\'an-Howarth formalism, the behavior of cross-scale energy transfer in magnetic reconnection and turbulence was found to be essentially
identical to each other, independent of an external magnetic (guide) field, in the inertial and energy-containing ranges (Adhikari et al., Phys. Plasmas 30, 082904, 2023). However, this description 
did not account for the energy transfer in the dissipation range for kinetic plasmas. In this letter, we adopt a scale-filtering approach to investigate this previously 
unaccounted-for energy transfer channel in reconnection. Using kinetic particle-in-cell (PIC) simulations of antiparallel and component reconnection, we show that the pressure-strain (PS) interaction becomes important at scales smaller than the ion inertial length, where the nonlinear energy transfer term drops off. Also, the presence of a guide field makes a significant difference in the morphology of the scale-filtered energy transfer. These results are consistent with kinetic turbulence simulations, suggesting that the pressure strain interaction is the dominant energy transfer channel between electron scales and ion scales. 

\end{abstract}

\section{Introduction}
A major theme of turbulence theory 
is the cascade of energy across scales, providing details of the succession of physical interactions that lead from large-scale energy input to small-scale dissipation and production of internal energy. This energy transfer 
\citep{Verma19}
is either studied in spectral space as a triadic interaction~\citep{domaradzki1990local,waleffe1992nature,ohkitani1992triad} or in lag space using the dynamical evolution of the second-order correlation tensors associated with the energy contained in the system~\citep{de1938statistical,monin1975statistical,politano1998dynamical}. The latter approach is quantified using the well-known von K\'arm\'an-Howarth (vKH) equation, which was applied primarily to hydrodynamic turbulence and in recent years has been generalized to account for plasma turbulence~\citep{galtier2008karman,banerjee2013exact, hellinger2018karman, andres2018energy}, including compressible MHD and its extensions~\citep{andres2018energy,banerjee2020scale,ferrand2021compact,simon2022exact}.

The vKH equation is a purely fluid construct. As such it lacks wave-particle interactions, separate contributions from ions and electrons, and other kinetic effects. We anticipate that the vKH equation based on fluid models remains credible for a kinetic plasma only at scales large enough to be well separated from kinetic effects namely the inertial and energy-containing ranges.
Using an incompressible vKH analysis of magnetic reconnection, we showed that the energy transfer characteristics in reconnection at MHD scales are qualitatively similar to that of a decaying turbulence across different scales~\citep{adhikari2021magnetic,adhikari2023effect}. However, one of the deficiencies
in the fluid description is the lack of a known form of dissipation for kinetic plasmas. As a result, a proper description of energy conversion and 
transfer at scales beyond the inertial range was missing. 

In addition to the 
description in terms of correlations, energy and energy flux distribution across different length scales can also be defined by a scale filtering operation. Recently, \citet{YangEA22} also showed that the scale-by-scale energy budget analysis using the vKH approach for a simulation of turbulence agrees with the scale-filtered energy equation resulting from the Vlasov-Maxwell equations in the inertial and energy-containing ranges. Moreover, the scale-filtered energy equation contains the scale-decomposed energy budget of the full Vlasov-Maxwell model. Therefore, the scale-filtering approach
can include 
the physics of energy transfer in Vlasov-Maxwell systems at scales not covered
in the vKH formalism,
thus providing a
more complete picture.

\section{Background}
Scale filtering~\citep{germano1992turbulence} is based on a properly defined filtering kernel $G_\ell = \ell^{-d}G(\mathbf{r}/\ell)$ which only maintains information about length scales $\geq \ell$. Here $G(\mathbf{r})$ is a non-negative normalized boxcar window function satisfying $\int d^d r G(\mathbf{r})=1$, and $d$ is the number of dimensions of the system. For any field $f(\mathbf{x},t)$, the scale-filtered field $\overline{f}_\ell(\mathbf{x},t)$ is defined as
\begin{equation}\label{eqn:scalefilter}
     \overline{f}_\ell(\mathbf{x},t) = \int d^d r G_\ell(\mathbf{r})f(\mathbf{x}+\mathbf{r},t).
\end{equation}

Likewise, the density-weighted filtered $f(\mathbf{x},t)$, also called the Favre-filtered field~\citep{favre1969statistical,aluie2013scale}, is defined as
\begin{equation}\label{eqn:favrefilter}
     \tilde{f}_\ell(\mathbf{x},t) = \frac{\overline{\left[\rho(\mathbf{x},t) f(\mathbf{x},t)\right]}_\ell}{\overline{\rho}_\ell(\mathbf{x},t)},
\end{equation}
where $\rho(\mathbf{x}.t)$ is the density. Following the scale filtering operation of the Vlasov-Maxwell equations, one can combine the electromagnetic energy, the total bulk flow energy, and the sub-grid scale energy flux to the pressure-strain interaction~\citep{yang2017energyb, matthaeus2020pathways,YangEA22} as
\begin{equation}
\label{eqn:scalefiltered}
     \underbrace{\partial_t \bigl< \sum_\alpha \Tilde{E}^f_\alpha \! + \overline{E}^m \bigr>}_{T_f-\epsilon} =  \underbrace{\! -\bigl< \sum_\alpha (\Pi^{uu}_\alpha \! + \Pi^{bb}_\alpha) \bigr>}_{-F_f}   \underbrace{\!  -\bigl< \sum_\alpha \Phi^{uT}_\alpha \bigr>}_{-D_f},
 \end{equation}
where $\alpha$ represents plasma species; $\Tilde{E}_\alpha^f = \frac{1}{2}\overline{\rho}_\alpha \Tilde{\mathbf{u}}_\alpha^2$ is the filtered bulk flow ($\mathbf{u}_\alpha$) energy density; $\overline{E}^m$ = $\frac{1}{8\pi} (\overline{\mathbf{B}}^2+\overline{\mathbf{E}}^2)$ is the filtered electromagnetic energy density; $\Pi_\alpha^{uu} = -(\overline{\rho}_\alpha \Tilde{\tau}^u_\alpha\cdot \nabla)\cdot \Tilde{\mathbf{u}}_\alpha - \frac{q_\alpha}{c}\overline{n}_\alpha\Tilde{\tau}_\alpha^b\cdot \Tilde{\mathbf{u}}_\alpha$ is the sub-grid scale flux term for bulk flow energy across scales due to nonlinearities, where $q_\alpha$ is the charge of plasma species $\alpha$, $\Tilde{\tau}^u_\alpha=\widetilde{\mathbf{u}_\alpha \mathbf{u}_\alpha}-\Tilde{\mathbf{u}}_\alpha \Tilde{\mathbf{u}}_\alpha$, $\Tilde{\tau}^b_\alpha=\widetilde{\mathbf{u}_\alpha\times \mathbf{B}}-\Tilde{\mathbf{u}}_\alpha\times \Tilde{\mathbf{B}}$. Similarly, $\Pi_\alpha^{bb}=-q_\alpha \Bar{n}_\alpha \Tilde{\mathbf{\tau}}_\alpha^e \cdot \Tilde{\mathbf{u}}_\alpha$ represents the sub-grid scale flux term for electromagnetic energy across scales due to nonlinearities, where $\Tilde{\mathbf{\tau}}_\alpha^e = \Tilde{\mathbf{E}}-\Bar{\mathbf{E}}$; $\Phi_\alpha^{uT}=-(\bar{\mathbf{P}}_\alpha\cdot \nabla)\cdot \Tilde{\mathbf{u}}_\alpha$ is the filtered pressure strain interaction that corresponds to the rate of conversion of flow into internal energy, where $\mathbf{P}_\alpha$ is the pressure tensor; and $\epsilon$ is the total dissipation rate.

In kinetic plasmas, where the exact form of dissipation may not be known, the total dissipation rate can be calculated by $\epsilon = -d\langle \sum_\alpha E_\alpha^f + E^m\rangle/dt=-\partial_t(\langle \sum_\alpha E_\alpha^f + E^m\rangle)-\mathbf{u} \cdot \nabla\langle \sum_\alpha E_\alpha^f + E^m\rangle.$ 

The first term in Eq.~\ref{eqn:scalefiltered} is the local time rate of change of flow and magnetic energy for scales $\geq \ell$. This term vanishes for large enough $\ell$ and approaches $-\epsilon$ as $\ell \xrightarrow{} 0$. With this property, one can define the time rate of change of energy for scales $< \ell$ as $T_f=\epsilon + \partial_t \langle \sum_\alpha \Tilde{E}^f_\alpha +\overline{E}^m \rangle$ such that $T_f\xrightarrow{}\epsilon$ at large $\ell$ and $T_f\xrightarrow{}0$ as $\ell \xrightarrow{}0.$ The first term on the right-hand side of Eqn.~\ref{eqn:scalefiltered} $F_f$ defined as $-\bigl< \sum_\alpha (\Pi^{uu}_\alpha \! + \Pi^{bb}_\alpha) \bigr>$, is associated with the nonlinear energy flux, while the second term  $D_f$, represents the internal energy deposition due to the pressure strain interaction, which is further decomposed as
\begin{equation}
    D_f = \sum_\alpha \langle -(\overline{\mathbf{P}}_\alpha\cdot \nabla) \cdot \Tilde{\mathbf{u}}_\alpha \rangle
    =\sum_\alpha \underbrace{\langle -\overline{p}_\alpha \nabla \cdot \Tilde{\mathbf{u}}_\alpha \rangle}_{\overline{p\theta_\alpha}} \underbrace{- \langle \overline{\mathbf{\Pi}}_\alpha : \Tilde{\mathbf{D}}_\alpha \rangle}_{\overline{PiD_\alpha}},
\end{equation}
where $p_\alpha=P_{\alpha,ii}/3$, $\Pi_{\alpha,ij}=P_{\alpha,ij}-p_{\alpha}\delta_{ij}$, and $D_{\alpha,ij}=(\partial_i u_{\alpha,j} + \partial_j u_{\alpha,i})/2-\nabla \cdot \mathbf{u}_{\alpha} \delta_{ij}/3$.
Finally, using these representations Eqn.~\ref{eqn:scalefiltered} can be rewritten in normalized form ($T_f^*=T_f/\epsilon$) as
\begin{equation}\label{eqn:symbolic}
    T_f^* + F_f^* + D_f^* = 1.
\end{equation}

For kinetic plasmas, Eqn.~\ref{eqn:symbolic} gives a generalized picture of energy transfer in several ways. On the one hand, unlike other equations based on MHD models (e.g., the vKH equation), Eqn.~\ref{eqn:symbolic} is entirely derived from the Vlasov-Maxwell model and therefore is an ideal candidate to describe its energy characteristics. On the other hand, under certain assumptions, Eqn.~\ref{eqn:symbolic} is analogous to the von K\'arm\'an-Howarth equation~\citep{de1938statistical,monin1975statistical},  derived through a completely different pathway using structure functions, based on increments. Both formalisms are expressions of conservation of energy across scales and are composed of different energy transfer terms.
One can find the correspondence between individual terms of Eqn.~\ref{eqn:symbolic} and the terms in the von K\'arm\'an-Howarth equation: $T_f^*$ is equivalent to the time rate of change of energy within a lag $\boldsymbol{l}$, $F_f^*$ is equivalent to the nonlinear energy transfer dominant in the inertial range and $D_f^*$ is equivalent to the visco-resistive dissipation in the MHD description. In a turbulence cascade scenario, different filtered terms are expected to dominate at different length scales: The time derivative term $T_f^*$ reaches the total dissipation rate $\epsilon$ at scales larger than the correlation length, decreases at intermediate scales (roughly the inertial range) where $F_f^*$ should dominate. At the smallest scales, the rate of production of internal energy $D_f^*$ (defined here as dissipation) becomes dominant.

In this paper, we use the scale-filtering approach to investigate the Vlasov-Maxwell picture of scale-to-scale energy transfer in magnetic reconnection using $2.5$D kinetic particle-in-cell (PIC) simulations. We find that the overall behavior of the scale-filtered energy equation in reconnection is similar to that of standard decaying turbulence in all the kinetic, inertial, and energy-containing ranges, with a better separation of scales with an increased guide field. In these systems, the filtered pressure strain interaction accounts for the energy transfer between the ion and electron scales. The remainder of this letter is organized as follows: In section~\ref{sec:simulation}, we provide the details of the reconnection simulations, followed by the results in section~\ref{sec:results}. Finally, section~\ref{sec:conclusion} summarizes the conclusions and discussions.

\section{Simulations} \label{sec:simulation}
In this study, we use five 2.5D kinetic particle-in-cell (PIC) simulations of magnetic reconnection performed using the P3D code~\citep{zeiler2002three}. These systems only vary by the magnitude of the out-of-plane magnetic (guide) field  $B_g$, which is normalized to the reference magnetic field $B_0$. The number density is normalized to $n_0$ while length scales are normalized to 
the ion inertial length $d_i = \sqrt{m_i c^2/4\pi n_0 e^2}$, time is normalized to the inverse of ion cyclotron frequency $\omega_{ci}^{-1} = (eB_0/m_i c)^{-1}$, and speed is normalized to the ion Alfv\'en speed $v_A=\sqrt{B_0^2/4\pi m_i n_0}$. Similarly, the electric field is normalized to $E_0=v_A B_0/c$, and temperature is normalized to $T_0=\frac{1}{2}m_iv_A^2$.

All the simulations are initialized with a double Harris equilibrium over a periodic square domain of size $L=204.8d_i$, and a sinusoidal perturbation of amplitude $0.12$ is added to initiate reconnection. These systems have a grid spacing of $\delta x=0.05d_i$ with $4096^2$ grid points, a time step of $\delta t\omega_{ci}=0.01$, and $100$ particles per cell with a total of $1.68\times 10^9$ particles per species. The mass ratio $m_i/m_e =25$, the speed of light $c=15v_A$, and the electron and ion temperature are initially set to $T_e=0.25,$ and $T_i=1.25$. The reconnecting field is set to $1B_0$ while the out-of-plane guide field is chosen from $[0,0.1,0.5,1,2]B_0$. The background density is set to $n_b=0.2$.

\begin{figure}
\centering
\includegraphics[scale=0.6]{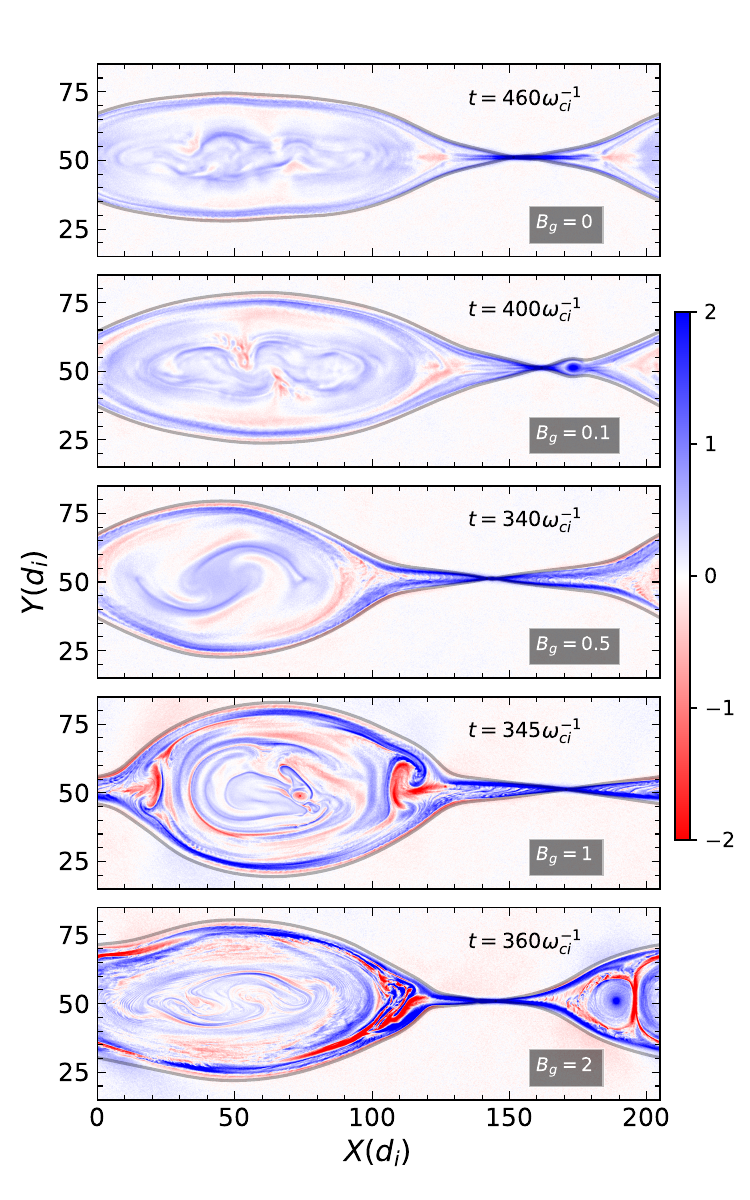}
\caption{\label{fig:vez}Overview of the reconnection simulations: out of plane electron velocity $v_{ez}$ in the lower current sheet for each guide field case along with the separatrix i.e. the contour of the magnetic flux function $\psi$ at the X-line. The time of analysis $t$ is given in the top right of each panel while the strength of the guide field $B_g$ is given on the bottom right.}
\end{figure}

Figure~\ref{fig:vez} provides an overview of all five simulations, showing the out-of-plane electron velocity $v_{ez}$ for the lower current sheet. Well-developed magnetic islands are visible in each run with different dynamics inside it. At the time of analysis, shown on the right top of each panel in Fig.~\ref{fig:vez}, all the runs have an equal amount of reconnected flux ($\Delta \psi\approx 15$) with slight variation in the island widths. With a larger guide field, the exhaust velocity increases~\citep{haggerty2018reduction}, allowing for a faster onset time. Note that the total energy is well-conserved in all of the simulations. For more details on the simulations, please refer to~\cite{adhikari2020reconnection,adhikari2021magnetic,adhikari2022reconnection}. Next, we discuss the scale-filtered energy analysis in reconnection.
\begin{figure}
\includegraphics[scale=0.65]{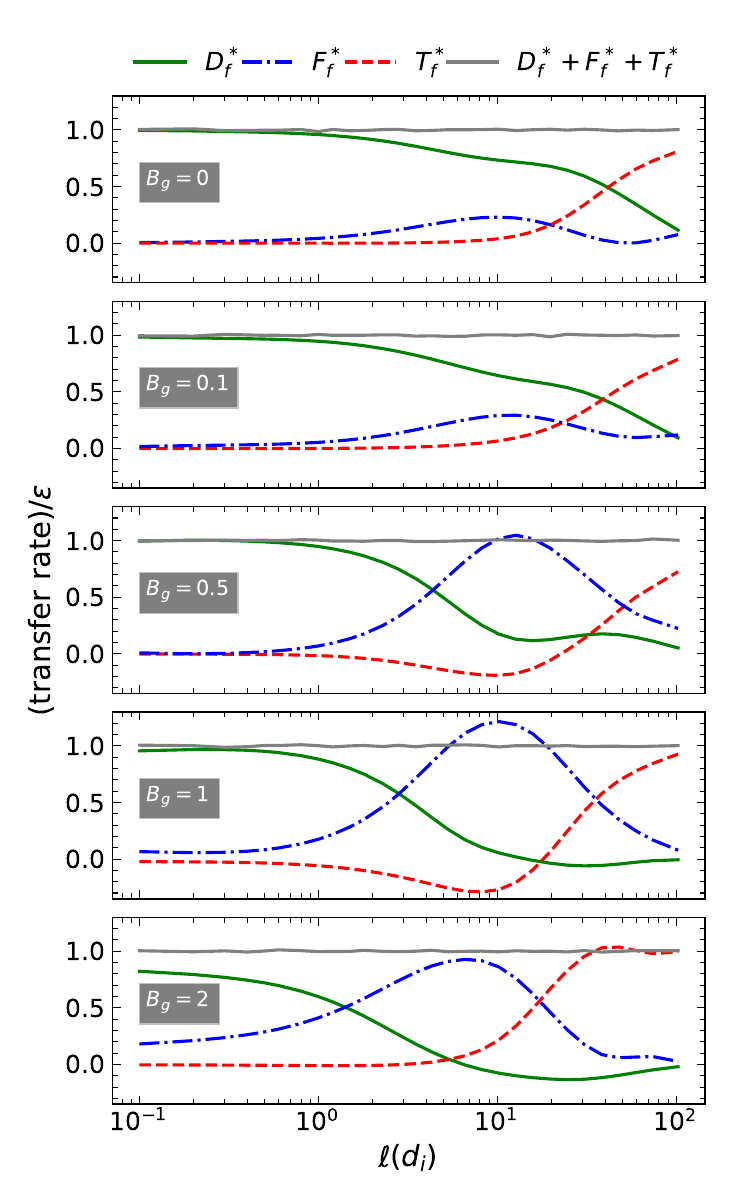}
\caption{\label{fig:scale_filter_energy}Individual terms of the scale-filtered energy equation (Eqn.~\ref{eqn:symbolic}) as a function of length scales $\ell$ for all the reconnection simulations. Each term is normalized to the value of $\epsilon=-d\langle \sum_\alpha E_\alpha^f +E_m \rangle/dt$ obtained from the simulation. The strength of the guide field used in the simulations is given on the left of each panel.}
\end{figure}
\section{Results}
\label{sec:results}
In Fig.~\ref{fig:scale_filter_energy}, we plot the individual terms of the scale-filtered energy equation (Eqn.~\ref{eqn:symbolic}) as a function of lag scale. The terms in Eqn.~\ref{eqn:symbolic} are calculated for a total of $3$ time slices centered at the time of analysis (shown in Fig.~\ref{fig:vez}), separated by $\Delta t=10\omega^{-1}_{ci}$ and the results shown are time-averaged over those three slices. 

In all cases studied here, the time derivative term $T_f^*$ is close to the total dissipation rate $\epsilon$ at very large scales, and decreases at intermediate scales. While the filtered pressure-strain term $D^*_f$ dominates over the nonlinear energy transfer term $F_f^*$ across all scales for smaller guide field runs ($B_g=0,0.1$), $F_f^*$ takes over $D_f^*$ in the inertial scales as the guide field becomes comparable or larger than the reconnecting magnetic field. However, the dominance of $D^*_f$ in the smaller scales still holds. This description of energy transfer at the smaller scales was previously missing in the vKH analysis of reconnection. If one compares this behavior of energy transfer with magnetohydrodynamic (MHD) turbulence, one might argue that $D^*_f$ is identical to the closed form of dissipation in kinetic plasmas. However, since our smallest filtering window is half of the electron inertial length $d_e/2$, we believe we need a much better resolution to understand the dissipation mechanisms at scales smaller than $d_e$. In any case, for the scales shown, the total energy conversion sums very closely to unity with only a slight fluctuation ($\leq 3\%$), thus accounting for all energy and providing a more complete view of 
the energy transfer
process.

\begin{figure}
\centering
\includegraphics[scale=0.65]{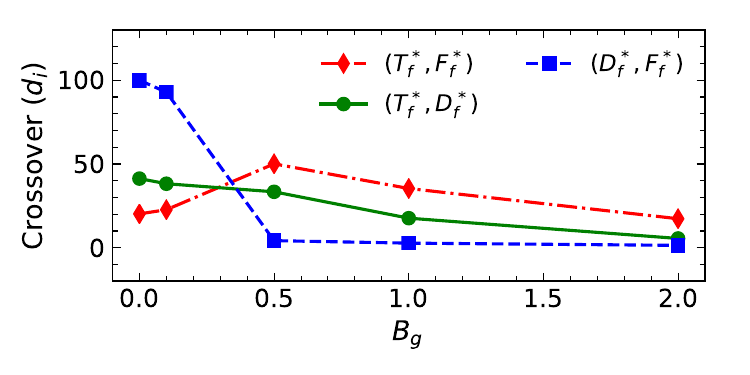}
\caption{\label{fig:crossover} A comparison of the crossover of the different pairs in Eqn.~\ref{eqn:symbolic} as a function of the guide field $B_g$.}
\end{figure}

The dominance of $F_f^*$ in the inertial scales with increased guide field is a consequence of a better separation of scales. That is, a well-separated inertial range exists, over which the dynamics is dominated by inertia terms while the time derivative term and the dissipation term become less prominent and almost negligible. For larger guide fields, the gyroradius of plasma species decreases, confining the electrons closer to the $X$ line. As a result, MHD physics dominates over a larger range extending to smaller scales. This behavior is illustrated in Fig.~\ref{fig:crossover}. For the larger guide field cases ($B_g\geq 0.5$), the point of crossover between the scale filtered pressure-strain (PS) interaction $D_f^*$ and nonlinear energy transfer term $F_f^*$ shifts sharply towards the smaller scales allowing the $F_f^*$ to dominate the intermediate scales. The crossover between the $D_f^*$ and $T_f^*$ terms also follow similar trend, however, the shift is relatively less drastic. On the other hand, the crossover between $F_f^*$ and $T_f^*$ terms has a different behavior, which shifts to larger scales until the guide field reaches $B_g=0.5$ and then falls back to the smaller scales. 



\begin{figure*}
\centering
\includegraphics[scale=0.57]{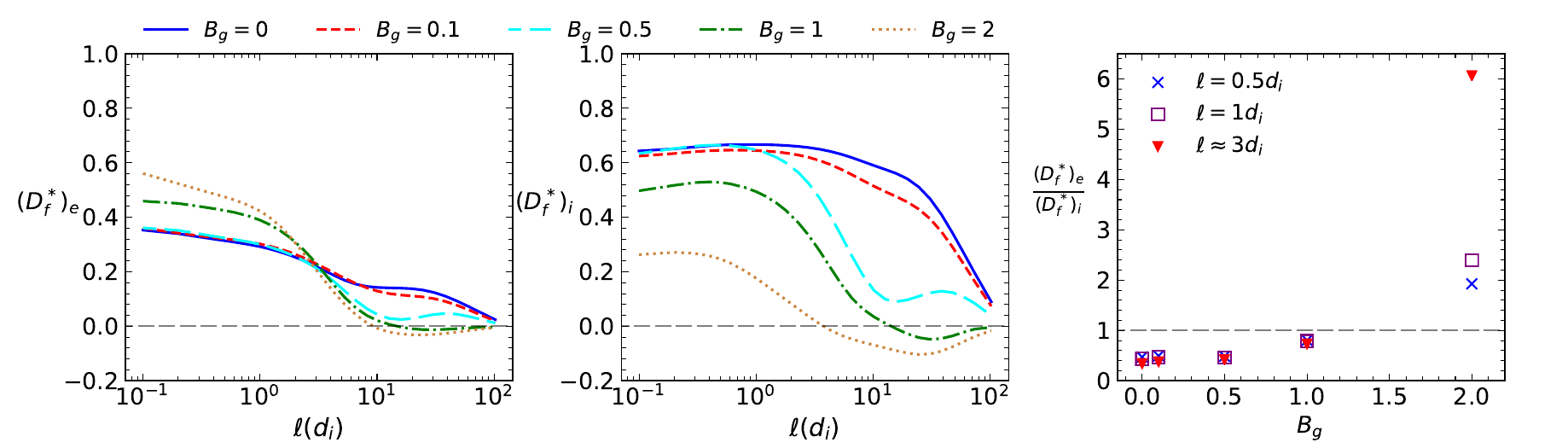}
\caption{\label{fig:ps_electron_ion} Scale filtering analysis of the pressure-strain interaction for electrons (left) and ions (middle) in all the simulations. The right column shows the scaling of the ratio of the filtered pressure strain interaction of electrons to ions with guide field for  lag $\ell= 0.5$, $1$, and $\approx 3d_i$.}
\end{figure*}

Next, we investigate the pressure-strain interactions of individual species. In Fig.~\ref{fig:ps_electron_ion}, we show the scale-filtered PS interaction $D_f^*$ for electrons (left column), ions (middle column), and the ratio between these two at different lag values (right column). It is observed that at smaller lag scales, the $D_f^*$ term for electrons is smaller than ions for smaller guide field as seen in the leftmost plot in Fig.~\ref{fig:ps_electron_ion}. However, for the largest guide field case, $(D_f^*)_e$ dominates over $(D_f^*)_i$. At larger scales, $D_f^*$ for both electrons and ions decreases with increasing guide field. This is certainly due to the role played by pressure dilatation ($\overline{p\theta}$) at  those scales. An increase in guide field decreases the compressibility in the system, which reduces the contribution of $\overline{p\theta}$. It is worth mentioning that for electrons, there is a crossover of $D_f^*$ from all the guide fields cases at $\ell \approx 2-3d_i$, while for ions no such crossover is observed. 
Finally, looking into the contributions to $D_f^*$, at smaller scales $D_f^*$ is dominated by $\overline{PiD}$ for both ions and electrons (not shown), an effect that becomes more pronounced with increasing guide field. Both incompressive and compressive pressure-strain interactions ($\overline{PiD}$, $\overline{p\theta}$) are stronger for ions compared to electrons at all scales for small guide field cases. This is the opposite of the largest guide field case. A detailed study on this effect is deferred to a later time. 

In the right most panel of Fig.~\ref{fig:ps_electron_ion}, we compare the ratio of the filtered PS term of electrons to ions for different lag scales of $0.5, 1,$ and $\approx 3d_i$. This ratio, a proxy of relative heating favours proton heating for $B_g \leq 1$, although a slight increase is seen in that range. However, for the stronger guide field $(B_g=2)$ electron heating is clearly favoured.

\section{Discussions and Conclusions}
\label{sec:conclusion}
In this paper, we study the behavior of the scale filtered energy equation in kinetic particle-in-cell simulations of both antiparallel and component reconnection. A guide field analysis of the filtered energy transfer equation reveal that for smaller guide fields, the pressure strain term dominates the non-linear energy transfer term across all scales. This is likely due to the limited system size affording insufficient scale separation that is required for 
strong turbulence and a well defined inertial range. However, for guide field larger or comparable to the reconnecting field, the pressure strain term becomes 
reduced in the inertial range, where the nonlinear energy transfer term dominates. This behavior is consistent to previous findings that the turbulence-like properties of reconnection 
are more pronounced with an external magnetic field~\citep{adhikari2023effect}. A
property worth mentioning is that the nonlinear term ($F_f^*$) peaks near the full values of the system dissipation rate $\epsilon$
at scales $\sim 10d_i$ in all the runs, suggesting a fully developed turbulence cascade rate, even if only over a small range of lags. 

In general, we note that he overall characteristics of the energy transfer equation are found similar to that of a decaying turbulence. This implies that the energy transfer in reconnection indeed resembles to that of turbulence at all scales and reconnection inherently involves an energy cascade.

The scale filtered analysis of the energy-budget equation has an advantage over the vKH formalism as it can account for the missing channel of energy transfer and conversion into internal energy at smaller scales that is represented by 
the filtered pressure-strain interaction~\citep{YangEA22}.

Present models that employ the vKH formalism assume fluid closures and therefore ignore the higher order physics involving the pressure tensor and its associated kinetic effects~\citep{DelSartoPegoraro16}.
On the other hand, the scale filtering analysis is based on the Vlasov equation and hence contains the self-consistent description of the production of internal energy (which we designate as ``dissipation'') 
via the pressure-strain term. However, one might be skeptical since the lowest filtering scales discussed in the numerical simulations are $\approx d_e$. A much better-resolved simulation is therefore needed to explore how the pressure-strain term or other physics would behave below electron length scales. This will shed light on whether the pressure-strain term can account for energy conversion at much smaller scales or some other kinetic physics needs to be explored. For example it is almost certain that the strongest guide field case must involve sub electron inertial scale 
physics based on the lack of convergence of 
pressure-strain at the smallest scale shown in Fig.~\ref{fig:scale_filter_energy}. 



\section*{Acknowledgments}
\begin{acknowledgments}
We thank the high-performance computing support from Cheyenne~\citep{Cheyenne18} provided by NCAR's Computational and Informations Systems Laboratory, sponsored by the NSF.
We also thank NERSC resources, a U.S. DOE Office of Science User Facility operated under Contract No. DE-AC02-05CH11231. 
S.~A. and P.~A.~C. are supported by DOE grant DE-SC0020294. Y.~Y 
and W.~H.~M are supported in part by NSF-DOE grant AGS 2108834,
by NASA under the MMS Theory, Modeling and Data Analysis project 80NSSC19K0565,
and by the LWS project under 
University of Maryland subcontract 98016-Z6338201.
M.~A.~S acknowledges support from NASA LWS grant 80NSSC20K0198.
\end{acknowledgments}

{}
\bibliographystyle{aasjournal}

\end{document}